\documentclass[12pt]{article}
\begin{document}
\begin{center}
{\bf{STUDY ON NONCOMMUTATIVE REPRESENTATIONS OF GALILEAN GENERATORS }}
\end{center}

\begin{center}
Sarmistha Kumar (Chaudhuri)$^{1}$, Saurav Samanta$^{2}$ \vskip .5cm
{$^1$ Camellia Institute of Technology, Madhyamgram, Kolkata-700129, India \\and\\S. N. Bose National Centre for Basic Sciences,
      JD Block, Sector III, \\Salt Lake, Kolkata-700098, India}\\      
{$^2$ Narasinha Dutt College,129, Belilious Road, Howrah-711101, India}\\{e-mail: srvsmnt@gmail.com}
\end{center}
\begin{abstract}
The representations of Galilean generators are constructed on a space where both position and momentum coordinates are noncommutating operators. A dynamical model invariant under noncommutative phase space transformations is constructed. The Dirac brackets of this model reproduce the original noncommutative algebra. Also, the generators in terms of noncommutative phase space variables are abstracted from this model in a consistent manner. Finally, the role of Jacobi identities is emphasised to produce the noncommuting structure that occurs when an electron is subjected to a constant magnetic field and Berry curvature. 
\end{abstract}

\section{Introduction}
It is generally believed that the measurement of space time coordinates at small scale involves unavoidable effects of quantum gravity. This effect, as suggested in the work of Doplicher {\it et. al.}\cite{doplicher,doplicher1}, can be incorporated in a physical theory by making the space time coordinates noncommutative. Without going into any detail one can write a general commutator among the space time coordinates as,
\begin{eqnarray}
[\hat{y}_{\mu}, \hat{y}_{\nu}]=i\theta_{\mu\nu}(\hat{y},\hat{q})\label{non}
\end{eqnarray}
Here $y,q$ are phase space variables.  The studies which are built on a structure like (\ref{non}) are called noncommutative physics\cite{review}. In the simplest nontrivial case one takes the noncommutative parameter $\Theta(=\theta_{\mu\nu})$ to be a constant antisymmetric matrix which is commonly named as canonical noncommutativity. Even in that case the commutator relation (\ref{non}) violates the Poincare symmetries\cite{kuldeep}.

In the last few years an interesting study has been found\cite{wess,wess1,wess2} where appropriate deformations of the representations of Poincare generators lead to different symmetry transformations which leave the basic commutator algebra covariant. In this way original Poincare algebra is preserved at the expense of modified coproduct rules. Quantum group theoretic approach following from the twist functions also give the identical results\cite{chaichian,chaichian1}. Construction of field theory based on these ideas and their possible consequences in field theory have been discussed in \cite{kuldeep,gonera}.

In nonrelativistic quantum mechanics, unlike space coordinates time is treated as a parameter instead of an operator. In that case though $\theta_{0i}=0$, remaining nonvanishing $\theta_{ij}$ breaks the Galilean invariance even for the canonical (constant $\theta$) case. But once again modifying the representations of generators one can keep the theory consistent with the noncommutating algebra among space coordinates. This has been shown in \cite{RB1} for the larger Schroedinger group, a subgroup of which is the Galilean group.

However, in all these analysis, the basic noncommutative brackets taken were somewhat restricted in the sense that noncommutativity among momenta coordinates were always taken to be zero. Interestingly, in the planar Landau problem which is frequently referred for the physical realization of canonical noncommutativity, it was shown in \cite{RB2,SS1} that noncommutativity among position coordinates and momenta coordinates has a dual nature. In the semiclassical treatment of Bloch electrons under magnetic field, a nonzero Berry curvature leads to a modification of the commutator algebra\cite{xiao}. When both the magnetic field and Berry curvature are constant the commutator brackets take a simple form and even in that case none of them is zero. On top of it even the standard position-momentum ($x-p$) algebra gets modified. In the present article we consider both position position and momentum momentum noncommutativity in 2+1 dimension and study the invariance of Galilean group. Before discussing further let us mention the summary of this paper.

 In the second section we give a general 
mapping between the commutating (which satisfy Heisenberg algebra) and noncommutating phase space variables. By a systematic method the values of different coefficients in this map are fixed. An inverse mapping is then obtained. In section 3, starting from a general noncommutative phase space algebra, we show how Jacobi identities lead to different brackets studied in earlier papers. Using the inverse map found in section 2, the appropriate noncommutative representations of the generators of the  Galilean group are obtained in the next section. These generators satisfy the usual closure algebra on the noncommutative plane. In section 5, using the realization of each generator
we calculate the symmetry transformations of the phase space coordinates. A dynamical model is then proposed in section 6. Constraint analysis of this model leads to nonzero Dirac brackets among position coordinates as well as momenta coordinates. These bracket structures are classical analogues of the quantum commutators considered in the earlier sections. Noether analysis is performed in the next section for the same model to get the classical version of the Galilean generators in terms of noncommutative phase space variables. Finally we conclude in section 8.

\section{Noncommutative Phase--Space}
In this section we show how  noncommutativity  can be introduced by suitable mapping
 of phase space variables obeying the commutative algebra. We have the standerd Heisenberg algebra
 in ( D=2+1 ) dimensional space as,
\begin{eqnarray}
&&[\hat{x}_i,\hat{x}_j]=0\nonumber\\
&&[\hat{p}_i,\hat{p}_j]=0\label{e1}\\
&&[\hat{x}_i,\hat{p}_j]=i\hbar\delta_{ij}\quad\quad(i=1,2)\nonumber
\end{eqnarray}
Here a quantum mechanical operator ($\hat{O}$) is denoted by putting a hat on its classical counterpart ($O$). Now we define two sets of variables $\hat{y}_i$ and $\hat{q}_i$ in terms of the commutative phase space variables in the following way
\begin{equation}
\hat{y}_i=\hat{x}_i+\alpha_1\epsilon_{ij}\hat{p}_j+\alpha_2\epsilon_{ij}\hat{x}_j\label{e2}
\end{equation}
\begin{equation}
\hat{q}_i=\hat{p}_i+\beta_1\epsilon_{ij}\hat{x}_j+\beta_2\epsilon_{ij}\hat{p}_j\label{e3}
\end{equation}
where $\alpha$ ($\alpha_1$, $\alpha_2$) and $\beta$ ($\beta_1$, $\beta_2$) are some arbitrary constants. Since for small values of $\alpha$ and $\beta$, ($\hat{y}, \hat{q}$) reduces to ($\hat{x}, \hat{p}$) we interpret $\hat{y}$ and $\hat{q}$ as modified coordinates and momomenta. Making use of (\ref{e1}) one finds that the new phase--space variables defined in the above two equations satisfy the algebra
\begin{equation}
[\hat{y}_i,\hat{y}_j]=-2i\hbar\alpha_1\epsilon_{ij}\label{eq4}
\end{equation}
\begin{equation}
[\hat{q}_i,\hat{q}_j]=2i\hbar\beta_1\epsilon_{ij}\label{e5}
\end{equation}
\begin{equation}
[\hat{y}_i,\hat{q}_j]=i\hbar(1+\alpha_2\beta_2-\alpha_1\beta_1)\delta_{ij}\label{eq6}
\end{equation}
Evidently the new brackets show the noncommutative nature of newly defined coordinates ($\hat{y}$) and momenta ($\hat{q}$). Henceforth they will be called noncommutative phase--space variables.
Note that a certain amount of flexibility is there due to different values of the constants $\alpha$ and $\beta$. We keep the bracket (\ref{eq6}) to its simplest undeformed form (\ref{e1}). This gives the condition
\begin{eqnarray}
\alpha_2\beta_2=\alpha_1\beta_1
\end{eqnarray}
Now without any loss of generality we can take $\alpha_2=\beta_2$ which fixes the constants $\alpha_2$ and $\beta_2$ in terms of the other two constants
\begin{eqnarray}
\alpha_2=\beta_2=\sqrt{\alpha_1\beta_1}\label{n1}
\end{eqnarray}
Next to give (\ref{eq4}) and (\ref{e5}) a neat form we set the following values of $\alpha_1$ and $\beta_1$
\begin{eqnarray}
&&\alpha_1=-\frac{\theta}{2}\label{n2}\\
&&\beta_1=\frac{\eta}{2}\label{n3}
\end{eqnarray}
where $\theta$ and $\eta$ are noncommutative parameters which in the present study are assumed to be nonzero. The choice of constants (\ref{n2}), (\ref{n3}) together with (\ref{n1}) yield the required noncommutative algebra
\begin{eqnarray}
&&[\hat{y}_i,\hat{y}_j]=i\hbar\theta\epsilon_{ij}\nonumber\\
&&[\hat{q}_i,\hat{q}_j]=i\hbar\eta\epsilon_{ij}\nonumber\\
&&[\hat{y}_i,\hat{q}_j]=i\hbar\delta_{ij}\label{eq8}
\end{eqnarray}
Such noncommutative structures appears in the chiral oscillator problem and the Landau model 
where a charged particle moves on a plane subjected to a strong perpendicular magnetic field. Phenomenological discussion of this structure was given in \cite{bertolami,bertolami1,bertolami2}.
 The inverse phase--space transformations of (\ref{eq8}) is given by,
\begin{eqnarray}
&&\hat{x}_i=A\hat{y}_i+B\epsilon_{ij}\hat{y}_j+C\hat{q}_i+D\epsilon_{ij}\hat{q}_j\nonumber\\
&&\hat{p}_i=E\hat{y}_i+F\epsilon_{ij}\hat{y}_j+A \hat{q}_i+B\epsilon_{ij}\hat{q}_j\label{eq9}
\end{eqnarray}
Here the various constants are,
\pagebreak
\begin{eqnarray*}
&&A=\frac{2-\theta\eta}{2(1-\theta\eta)}\quad\quad\quad
B=\frac{-\sqrt{-\theta\eta}}{2(1-\theta\eta)}\\
&&C=\frac{\theta\sqrt{-\theta\eta}}{2(1-\theta\eta)}\quad\quad\quad
D=\frac{\theta}{2(1-\theta\eta)}\\
&&E=\frac{-\eta\sqrt{-\theta\eta}}{2(1-\theta\eta)}\quad\quad\quad
F=\frac{-\eta}{2(1-\theta\eta)}
\end{eqnarray*}
Note that the hermiticity of physical operators $x, p$ and 
$\hat{y}$ , $\hat{q}$ can be restored by demanding different signs of $\theta$ and $\eta$ which will keep the various co-efficients  real and well defined.
\section{Role of Jacobi Identities in Planar Noncommutativity }
Jacobi identities are known to play an important role in fixing the structure of the noncommutative algebra. For instance in \cite{magg} the algebra of Kappa-deformed space was obtained in this manner. In this section we discuss the obtaining of planar noncommutative algebra by exploiting Jacobi identities.

Consider a plane where the noncommutative parameters are not constants. They are taken to be arbitrary functions of the position coordinates. Since Jacobi identities must be satisfied for the phase space commutator algebra, the functions appearing in the brackets cannot all be independent. The relations among these functions will enable us to generate different types of noncommutative structures studied in earlier papers. 

We take the noncommutative structure in the following form
\begin{eqnarray}
&&[\hat{y}_i,\hat{y}_j]=i\hbar\Omega f(x)\epsilon_{ij}\label{1001}\\
&&[\hat{q}_i,\hat{q}_j]=i\hbar Bg(x)\epsilon_{ij}\label{1002}\\
&&[\hat{y}_i,\hat{q}_j]=i\hbar s(x)\delta_{ij}\label{1003}
\end{eqnarray}
where $\Omega, \ B$ are constants and $f,g,s$ are some functions of coordinates.
The Jacobi identity for $y_i$--$q_j$--$q_k$ is
\begin{eqnarray}
[y_i,[q_j,q_k]]+[q_j,[q_k,y_i]]+[q_k,[y_i,q_j]]=0
\end{eqnarray}
Using (\ref{1002}) and (\ref{1003}) in the above equation we find
\begin{eqnarray}
(B\Omega f\partial_kg-s\partial_k s)\delta_{ij}-(B\Omega f\partial_jg-s\partial_js)\delta_{ik}=0\label{1004}
\end{eqnarray}
where $\partial_k=\frac{\partial}{\partial x_k}$ in the above equation. Similarly the Jacobi identitity for  $y_i$--$y_j$--$q_k$ give
\begin{eqnarray}
&&f\partial_ms(\epsilon_{im}\delta_{jk}-\epsilon_{jm}\delta_{ki})-s\partial_kf\epsilon_{ij}=0\label{1006}
\end{eqnarray}
Other two Jacobi identities are identically zero. In the equations (\ref{1004}) and (\ref{1006}), $i,j,k$ take values only 1 and 2. Now we take specific choices for $i,j,k$ in these equations to get simplified equations. For $i=1,j=2,k=2$ (\ref{1006}) gives
\begin{eqnarray}
f\partial_2s-s\partial_2f=0\label{001}
\end{eqnarray}
And the choice $i=2,j=1,k=1$ in the same equation gives
\begin{eqnarray}
f\partial_1s-s\partial_1f=0\label{001a}
\end{eqnarray}
Equations (\ref{001}) and (\ref{001a}) are written in a covariant way
\begin{eqnarray}
f\partial_is-s\partial_if=0\label{AA}
\end{eqnarray}
Equation (\ref{1004}) under the choices $i=1,j=1,k=2$ and  $i=2,j=1,k=2$ gives following two equations
\begin{eqnarray}
&&B\Omega f\partial_2g-s\partial_2s=0\label{002}\\
&&B\Omega f\partial_1g-s\partial_1s=0\label{003}
\end{eqnarray}
These two equations are written in a covariant manner
\begin{eqnarray}
B\Omega f\partial_ig-s\partial_is=0\label{BB}
\end{eqnarray}
Equation (\ref{AA}) immediately implies
\begin{eqnarray}
\partial_i\big(\frac{s}{f}\big)=0
\end{eqnarray}
Or equivalently,
\begin{eqnarray}
s(x)=\xi f(x)\label{CC}
\end{eqnarray}
where $\xi$ is some arbitrary constant. Replacing $f$ in terms of $s$ in the commutator algebra (\ref{1001}), we see the constant $\frac{1}{\xi}$ can be absorbed in $\Omega$. So without any loss of generality we set $\xi=1$ in (\ref{CC}) to get
\begin{eqnarray}
f(x)=s(x)\label{DD}
\end{eqnarray}
Substituting (\ref{DD}) in (\ref{BB}) we find
\begin{eqnarray}
s\partial_i(B\Omega g-s)=0
\end{eqnarray}
Thus the term within the parentheses is a constant and we write it as $-\lambda$ {\it i.e.}
\begin{eqnarray}
g(x)=\frac{1}{B\Omega}(s(x)-\lambda)\label{EE}
\end{eqnarray}
Equations (\ref{DD}) and (\ref{EE}) give severe restrictions on the structure of the noncommutative algebra (\ref{1001})--(\ref{1003}). For example, if we set $s=1$, then from (\ref{DD}) and (\ref{EE}) we get
\begin{eqnarray}
&&f=1\\
&&g=\frac{1-\lambda}{B\Omega}
\end{eqnarray}
These when substituted in (\ref{1001})--(\ref{1003}) give
\begin{eqnarray}
&&[\hat{y}_i,\hat{y}_j]=i\hbar\Omega\epsilon_{ij}\label{1001'}\\
&&[\hat{q}_i,\hat{q}_j]=i\hbar \frac{1-\lambda}{\Omega}\epsilon_{ij}\label{1002'}\\
&&[\hat{y}_i,\hat{q}_j]=i\hbar\delta_{ij}\label{1003'}
\end{eqnarray}
This noncommutative structure is nothing but the algebra (\ref{eq8}) under the identification $\Omega=\theta$ and $\frac{1-\lambda}{\Omega}=\eta$.

It is interesting to take a different choice of $s$
\begin{eqnarray}
s(x)=g(x)
\end{eqnarray}
Then from (\ref{EE})
\begin{eqnarray}
g=\frac{\lambda}{1-B\Omega}\label{FF}
\end{eqnarray}
Equations (\ref{DD}), (\ref{EE}) and (\ref{FF}) show that $f,g$ and $s$ are constant and
\begin{eqnarray}
f=g=s=\frac{\lambda}{1-B\Omega}\label{FF1}
\end{eqnarray}
Using the above equation in (\ref{1001})--(\ref{1003}) we get
\begin{eqnarray}
&&[\hat{y}_i,\hat{y}_j]=i\hbar\frac{\Omega}{1-B\Omega}\epsilon_{ij}\label{1001''}\\
&&[\hat{q}_i,\hat{q}_j]=i\hbar \frac{B}{1-B\Omega}\epsilon_{ij}\label{1002''}\\
&&[\hat{y}_i,\hat{q}_j]=i\hbar\frac{1}{1-B\Omega}\delta_{ij}\label{1003''}
\end{eqnarray}
where we set $\lambda=1$ which appeared as an overall scaling. Above noncommutative structure appears when an electron is subjected to a uniform magnetic field ($B$) and constant Berry curvature ($\Omega$)\cite{xiao}. Further discussion on this algebra may be found in \cite{Duval,Duval1,Duval2}.
\section{Realizations of Galilean Generators in Noncommutative Phase Space}
In this section we will study the representations of the generators of Galilean group in the noncommutative plane characterized by (\ref{eq8}){\footnote{Symmetry analysis for particular type of nonconstant noncommutative parameter may be found in \cite{RSS,Luki2}.}}.This group consists of  Angular momentum ($\hat{J}$),
  Translation ($\hat{\bf{P}}$), and Boost ($\hat{\bf{G}}$) which
 in 2+1 dimensional commutative space take the form
\begin{eqnarray}  
&&\hat{J}=\epsilon_{ij}\hat{x}_i\hat{p}_j\nonumber\\
&&\hat{P}_i=\hat{p}_i\nonumber\\
&&\hat{G}_i=m\hat{x}_i-t\hat{p}_i\label{eq10}
\end{eqnarray}
They are known to satisfy the following closure properties
 \begin{eqnarray}
 &&[\hat{P}_i,\hat{P}_j]=0\quad\quad\quad   [\hat{G}_i,\hat{J}]=-i\hbar\epsilon_{ij}\hat{G}_k\nonumber\\
&&[\hat{G}_i,\hat{G}_j]=0\quad\quad\quad[\hat{J},\hat{J}]=0\nonumber\\
 &&[\hat{P}_i,\hat{J}]=-i\hbar\epsilon_{ij}\hat{P}_k\quad\quad\quad
[\hat{P}_i,\hat{G}_j]=-im\hbar\delta_{ij}\label{eq11}
 \end{eqnarray}
 In a 2+1 dimensional noncommutative space (\ref{eq8}), mere substitution of old phase space variables ($x$, $p$) by 
new variables ($y, q$) would not preserve the closure algebra. Thus, it is necessary to change the standard representations of the generators in an appropriate manner so that the symmetry remains invariant. The new representations are readily obtained by using the mapping (\ref{eq9}) in the expressions (\ref{eq10}). These are given by,
\begin{eqnarray}
\hat{J}&=&\frac{E}{2}\epsilon_{ij}\hat{y}_i\hat{y}_j+\frac{C}{2}\epsilon_{ij}\hat{q}_i\hat{q}_j+
\frac{1}{1-\theta\eta}\epsilon_{ij}\hat{y}_i\hat{q}_j+D{\hat{q}_j}^2-F\hat{y}_j^2
\label{j}\\
\hat{P}_i&=&E\hat{y}_i+F\epsilon_{ij}\hat{y}_j+A \hat{q}_i+B\epsilon_{ij}\hat{q}_j
\label{pi}\\
\hat{G}_i&=&m(A\hat{y}_i+B\epsilon_{ij}\hat{y}_j+C\hat{q}_i+D\epsilon_{ij}\hat{q}_j)\nonumber\\
&&-t(E\hat{y}_i+F\epsilon_{ij}\hat{y}_j+A \hat{q}_i+B\epsilon_{ij}\hat{q}_j)
\label{eq12}
\end{eqnarray}
It can be easily verified that in the noncommutative space (\ref{eq8}) the generators with their new realizations (\ref{j}--\ref{eq12}) satisfy the undeformed algebra (\ref{eq11}). Quite naturally in the limit of vanishing noncommutative parameters ($\theta,\eta\rightarrow 0$) (\ref{j}--\ref{eq12}) reduce to the primitive form of the generators (\ref{eq10}) under the identification $(\hat{y},\hat{q})\rightarrow(\hat{x},\hat{p})$.
\section{Infinitesimal Symmetry Transformations}
The explicit presence of noncommutative parameters in the phase space algebra hints at possible 
changes in the symmetry transformations. These are obtained by calculating the commutator of noncommutative phase space variables and the generators with their new representations listed in (\ref{j}--\ref{eq12}). The results are given seperately for each generators. \\\\
{\bf Translation}\\
We first consider the translation generator ($P$) which gives the following transformation rule for the noncommutative coordinate $\hat{y}$
\begin{eqnarray}
\delta\hat{y}_i&=&\frac{i}{\hbar}[\hat{P},\hat{y}_i]\nonumber\\
&=&\frac{i}{\hbar}a_k[E\hat{y}_k+F\epsilon_{kj}\hat{y}_j+A\hat{q}_k+B\epsilon_{kj}\hat{q}_j,\hat{y}_i]\nonumber\\
&=&{a_i}+\frac{1}{2}\sqrt{-\theta\eta}\epsilon_{ij}a_j\label{100}.
\end{eqnarray}
The transformation rule for momentum variable is obtained in a likewise manner
\begin{equation}
\delta{\hat{q}_i}=\frac{\eta}{2}\epsilon_{ij}a_j\label{eq14.1}
\end{equation}
Expectedly, $(\theta,\eta)\rightarrow 0$ gives the correct commutative space results for the expressions (\ref{100}) and (\ref{eq14.1}).\\\\
{\bf Rotation}\\
An identical treatment for the rotation generator gives the following transformation rules for the phase space variables
\begin{eqnarray}
\delta\hat{y}_i&=&\frac{i}{\hbar}\alpha[\hat{J},\hat{y}_i]\nonumber\\
&=&-\alpha\epsilon_{ik}\hat{y}_k\nonumber\\
\delta{\hat{q}_i}&=&\frac{i}{\hbar}\alpha[\hat{J},\hat{q}_i]\nonumber\\
&=&-\alpha\epsilon_{ik}\hat{q}_k\label{eq14.2}
\end{eqnarray}
Interestingly, these expressions are identical with the corresponding transformation rules for the commutative space. This is a very special property which holds only in 2+1 dimension.\\\\
{\bf Boost}\\
Similarly for the boost generator the transformation rules are found to be
\begin{eqnarray}
\delta\hat{y}_i&=&\frac{m\theta}{2}\epsilon_{ik}a_k-t(\frac{\sqrt{-\theta\eta}}{2}\epsilon_{ik}a_k
+{a_i})\nonumber\\
\textrm{and}\quad\quad\delta\hat{q}_i&=&-m({a_i}+\frac{\sqrt{-\theta\eta}}{2}\epsilon_{il}a_l)-
t(\frac{\eta}{2}\epsilon_{il}a_l)\label{eq14.3}
\end{eqnarray}
These expressions also have the smooth commutative limit $(\theta,\eta)\rightarrow 0$.
\section{Construction of a Dynamical Model}
 In order to generate the noncommutative algebra (\ref{eq8}) in a natural way from a model, we consider the first order form of the non relativistic free particle Lagrangian 
\begin{equation}
L= p_i\dot{x}_i-\frac{p^2}{2m}\label{eq15}
\end{equation}
and use the classical version of the transformation (\ref{eq9}) to get the following form of the Lagrangian
\pagebreak
\begin{eqnarray}
L&=&[\frac{E}{2}y_k\dot{y}_k+F\epsilon_{kl}\dot{y}_ky_l+Mq_k\dot{y}_k
  +\frac{C}{2}q_k\dot{q}_k+D\epsilon_{kl}q_k\dot{q}_l]\nonumber\\
&&-\frac{1}{2m}[(E^2+F^2){y_i}^2+(A^2+B^2)q_i^2+
Ey_iq_i-2F\epsilon_{ij}y_iq_j]\label{eq16}
\end{eqnarray}
where $M=1/(1-\theta\eta)$ in the above equation. For $\eta=0$, physical application of this model was studied in \cite{Duval} and theoretical discussion, notably Lagrangian involving second order time derivative was given in \cite{look,Luki1}. The relation between Chern-Simons field theory and the Landau problem in the noncommutative plane has been studied in \cite{new}. In the above model we interpret $y$ and $q$ as the configuration space variables in an extended space.
The canonical momenta conjugate to $y$ and $q$ are,
\begin{eqnarray}
&&{\pi_i}^y=\frac{E}{2}y_i+F\epsilon_{ik}y_k+Mq_i\label{eq17a}\\
&&{\pi_i}^q=\frac{C}{2}q_i-D\epsilon_{ik}q_k\label{eq17}
\end{eqnarray}
These momenta ($\pi_i^y,\pi_i^q$) together with the configuration space variables ($y_i,q_i$) give the following Poisson algebra
\begin{eqnarray}
&&\{y_i,\pi_j^y\}=\delta_{ij}\\
&&\{q_i,\pi_j^q\}=\delta_{ij}
\end{eqnarray}
All other brackets are zero.
Since none of the momenta ((\ref{eq17a}) and (\ref{eq17})) involve velocities these are interpreted as primary constraints.
These are given by,
\begin{eqnarray}
\Omega_{1,i}&=&{\pi_i}^y-[\frac{E}{2}y_i+F\epsilon_{ik}y_k+Mq_i] \approx 0\nonumber\\
\Omega_{2,i}&=&{\pi_i}^q-[\frac{C}{2}q_i-D\epsilon_{ik}q_k]\approx 0\label{eq18}
\end{eqnarray}
They satisfy the Poisson algebra
\begin{eqnarray}
&&\{\Omega_{1,i},\Omega_{1,j}\}=-2F\epsilon_{ij}\nonumber\\
&&\{\Omega_{1,i},\Omega_{2,j}\}=-M\delta_{ij}=-\{\Omega_{2,i},\Omega_{1,j}\}\nonumber\\
&&\{\Omega_{2,i},\Omega_{2,j}\}=2D\epsilon_{ij}\label{eq19}
\end{eqnarray}
  Evidently $\Omega_{1,i}$ and $\Omega_{2,i}$ do not close among themselves so they are the second class constraints according to Dirac's classification\cite{Dirac}. This set can be eliminated by computing Dirac brackets. For that we write the constraint matrix. 
 \begin{eqnarray}
 \Lambda_{ij} =\big( \Lambda_{ij}^{mn}\big)&=&\left(\matrix{
{\{\Omega_{1,i},\Omega_{1,j}\}}&{\{\Omega_{1,i},\Omega_{2,j}\}}\cr
{\{\Omega_{2,i},\Omega_{1,j}\}} & {\{\Omega_{2,i},\Omega_{2,j}\}}}\right)\nonumber\\&=&\left(\matrix{
{-2F\epsilon_{ij}}&{-M\delta_{ij}}\cr
{M\delta_{ij}} & {2D\epsilon_{ij}}}\right)
\label{eq20}
\end{eqnarray}
We write the inverse of $\Lambda_{ij}$ as $\Lambda^{(-1)}_{ij}$ such that ${\Lambda_{ij}}^{mn}\Lambda_{jk}^{(-1)ns}=
{\delta^{ms}}_{ik}$ $(\textrm{for} \  i,j,k=1,2)$.
It is given by,
\begin{eqnarray}
\Lambda_{ij}^{(-1)ns}=\left(\matrix{
{\theta}\epsilon_{ij} & \delta_{ij}\cr
-\delta_{ij} & {\eta}\epsilon_{ij}}\right)
\label{eq21}
\end{eqnarray}
Using the definition of Dirac bracket\cite{Dirac}
\begin{displaymath}
\{f,g\}_{DB}=\{f,g\}-\{f,\Phi_{i,m}\}{\Lambda_{ij}}^{(-1)ms}\{\Phi_{j,s},g\}
\end{displaymath}
our model yields the following Dirac brackets in configurationa space 
\begin{eqnarray}
\{y_i,y_j\}&=&{\theta}\epsilon_{ij}\nonumber\\
\{q_i,q_j\}&=&{\eta}\epsilon_{ij}\nonumber\\
\{y_i,q_j\}&=&\delta_{ij}\label{eq22}
\end{eqnarray}
This algebra manifests the classical analog of the noncommutative algebra ({\ref{eq8}).
\section{ Noether's theorem and generators}
In this section we reproduce the noncommutative representations of Galilean generators from a Noether analysis of the Lagrangian ({\ref{eq16}}). The realizations of generators thus obtained from a knowledge of infinitesimal symmetry transformation will be shown to be identical with those found in section 4. This clearly shows the consistency between the dynamical approach of previous section and the algebraic approach of section 4.

The invariance of an action 
$S$ under an infinitesimal symmetry transformation
\begin{equation}
\delta{Q_i}=\{G,Q_i\}\nonumber\\
\label{eq23}
\end{equation}
is given by,
\begin{equation}
\delta{S} =\int{dt} \delta L=\int{dt}{\frac{d}{dt}}(\delta{Q_i}P_i-G)\label{eq24}   
\end{equation}
where G is the generator of the transformation and $P_i$ is the canonical momenta
conjugate to $Q_i$. If we denote the quantity inside the parentheses by $B(Q,P)$, then
\begin{equation}
G=\delta{Q_i}{P_i}-B\nonumber\\   
\end{equation}
Then this can be taken as the definition of the generator $G$. For the model (\ref{eq16}) both $y$ and $q$ have been interpreted as 
configuration space variables so we write the above equation as,
\begin{equation}
G=\delta{q_i}{\pi_i}^q+\delta{y_i}{\pi_i}^y-B\label{eq25} 
\end{equation}
Using the expressions of the canonical momenta (\ref{eq17a},\ref{eq17}), above equation is written in an explicit form as, 
\begin{eqnarray}
  G=\delta{y_i}\big(\frac{E}{2}y_i+F\epsilon_{ik}{y_k}+Mq_i\big)
+\delta{q_i}\big(\frac{C}{2}q_i-D\epsilon_{ik}q_k\big)-B
\label{eq26}
\end{eqnarray}
Knowing the transformation rules of $y$ and $q$, this relation is now used to find the representations of the generators one by one.\\ \\
{\bf{Translations}}\\
It is obvious from (\ref{100}) and (\ref{eq14.1}) that the time derivatives of the variations
$\delta{y}_i$ and $\delta{q}_i$ are zero. So from (\ref{eq16}) we write $\delta{L}$ as,
\begin{eqnarray}
\delta{L}&=&[\frac{E}{2}\delta{y}_k+F\epsilon_{kl}\delta{y}_l+M\delta{q}_k]\dot{y}_k+[\frac{C}{2}\delta{q}_k
+D\epsilon_{lk}\delta{q}_l]\dot{q}_k\nonumber\\
&&-\frac{1}{2m}[((E^2+F^2)2y_i+Eq_i-2F\epsilon_{ij}q_j)\delta{y}_i+\nonumber\\
&&((A^2+B^2)2q_i+Ey_i-2F\epsilon_{ki}y_k)\delta{q}_i]\nonumber\\
&&=\frac{d}{dt}[\frac{E}{4}\sqrt{-\theta\eta}\epsilon_{ks}y_{k}a_{s}+\frac{D\eta}{2}a_kq_k+\frac{C\eta}{4}\epsilon_{ks}q_ka_s]\label{B}
\end{eqnarray}
Using (\ref{B}) and the phase space transformation rules (\ref{100}, \ref{eq14.1}) in (\ref{eq25})
we get the translational generator,
\begin{eqnarray}
G_{Tr}=a_i(E\hat{y}_i+F\epsilon_{ij}\hat{y}_j+A \hat{q}_i+B\epsilon_{ij}\hat{q}_j)\label{eq27}
\end{eqnarray}
This result matches exactly with the expression of the translation generator obtained in (\ref{pi}).
\\\\
{\bf{Rotation}}\\
\vskip 0.05cm
Under rotation the transformations (\ref{eq14.2}) gives the following variation of the Lagrangian 
\begin{eqnarray*}
\delta{L}&=&[\frac{E}{2}(\delta{y}_k\dot{y}_k+y_k{\delta{\dot {y}}_k})+F\epsilon_{kl}({\delta{\dot{y}}}_ky_l
+y_l\dot{\delta{y}_k})\nonumber\\
&&+\frac{C}{2}(\delta{q}_k\dot{q}_k+q_k{\delta{\dot{q}}_k})+
D\epsilon_{kl}(\delta{q}_k\dot{q}_l+q_k{\delta{\dot{q}}_l})
+M(\delta{q}_k\dot{y}_k+q_k\delta{\dot{y}}_k)]\nonumber\\
&&-\frac{1}{2m}[(E^2+F^2)2y_i+Eq_i-2F\epsilon_{ij}q_j)\delta{y}_i\nonumber\\&&+
((A^2+B^2)2q_i+Ey_i-2F\epsilon_{ki}y_k)\delta{q}_i]\nonumber\\
&=&\frac{d}{dt}(0)\nonumber\\
\end{eqnarray*}

Since $B=0$ for the Lagrangian (\ref{eq16}) under rotation we obtain from (\ref{eq26}) and (\ref{eq14.2})
 the desired form of rotational generator 
\begin{equation}
G_{Rot}=\alpha({\frac{E}{2}\epsilon_{ij}\hat{y}_i\hat{y}_j+\frac{C}{2}\epsilon_{ij}\hat{q}_i\hat{q}_j+
\frac{1}{1-\theta\eta}\epsilon_{ij}\hat{y}_i\hat{q}_j+D{\hat{q}_j}^2-F\hat{y}_j^2}) 
\label{eq28}
\end{equation}
Here also the realization of rotational generator is same as (\ref{j}).
\\\\
{\bf{Boost}}\\
Following similar approach for the Lagrangian (\ref{eq16}) in case of Boost symmetry (\ref{eq14.3}) we find,
\begin{eqnarray*}
\delta{L}&=&\frac{d}{dt}[a_iy_i\{-m(M+\frac{\theta}{2}F\}+a_k\epsilon_{ki}y_im(-\frac{E\theta}{4}+\frac{M}{2}
\sqrt{-\theta\eta})\nonumber\\
&&+a_iq_i(-mC)+a_k\epsilon_{ki}q_im(\frac{C}{4}\sqrt{-\theta\eta}-D)]+
\frac{d}{dt}[a_i\frac{F\theta}{2}{q_i}t
+a_k\epsilon_{ki}\frac{\sqrt{-\theta\eta}}{2}\frac{E}{2}y_it\nonumber\\
&&+a_k\epsilon_{ki}q_i\frac{C\eta}{4}t]\nonumber
\end{eqnarray*}

Thus we can identify $B$ from the above result and using this in (\ref{eq26}) together with (\ref{eq14.3}) we achieve the expression for the Boost generator 
\begin{eqnarray}
G_{Boost}&=&a_i[m(A\hat{y}_i+B\epsilon_{ij}\hat{y}_j+C\hat{q}_i+D\epsilon_{ij}\hat{q}_j)\nonumber\\&&-
t(E\hat{y}_i+F\epsilon_{ij}\hat{y}_j+A \hat{q}_i+B\epsilon_{ij}\hat{q}_j)]\nonumber
\end{eqnarray}
Thus again the expression of Boost generator is identical with (\ref{eq12}) found from algebraic approach. 
\section{ Conclusions}
We have considered a plane where not only position variables but also momentum variables are intrinsically noncommutating. This is an important departure from earlier studies in this context where noncommutativity appeared only in position position coordinates. Imposing the Jacobi identities among the various variables we were able to reproduce, from general arguments, the specific structure of noncommutativity discussed in the context of physical models. We obtained the representations of Galilean generators in an algebraic approach which are compatible with this noncommutative space. We also constructed a dynamical model invariant under symmetry transformation rules of phase space variables. Constraint analysis of this model allowed us to identify the second class constraints which finally lead to the noncommutative Dirac brackets. These brackets are the classical analogs of the noncommutative algebra.  Finally Noether theorem was applied to this dynamical model to obtain the noncommutative representations of the generators. The realizations of these generators were identical with those found by the previous method. In this way consistency between the algebraic approach and the dynamical approach was established.
\vskip 1.5cm

{\bf Acknowledgment}\\

Both the authors thank Prof. R. Banerjee for suggesting this investigation. One of the authors, Sarmistha Kumar (Chaudhuri) would like to thank the Director, {\it S.N.Bose National Centre for Basic Sciences,Block-JD,Sector-III,Salt Lake} for providing the facilities to carry out this work.

\end{document}